\documentclass[twocolumn,twoside,slac_two]{revtex4}
\usepackage{graphicx}
\usepackage{fancyhdr}
\begin{document}

\title{High-energy gamma-ray observations of Geminga with the Fermi Large Area Telescope}

%

\author{M. Razzano}
\affiliation{Istituto Nazionale di Fisica Nucleare, Sezione di Pisa, I-56127 Pisa, Italy}
\author{D. Dumora}
\affiliation{Universit\'e de Bordeaux, Centre d'\'Etudes Nucl\'eaires Bordeaux Gradignan, UMR 5797, Gradignan, 33175, France}

\author{F. Gargano}
\affiliation{Istituto Nazionale di Fisica Nucleare, Sezione di Bari, 70126 Bari, Italy}
\author{\textit{on behalf of the Fermi Large Area Telescope Collaboration}}
\noaffiliation
\begin{abstract}
Geminga is the second brightest persistent source in the GeV $\gamma$-ray sky. Discovered in 1975 by SAS-2 mission, it was identified as a pulsar only in the 90s, when ROSAT detected the 237 ms X-ray periodicity, that was later also found by EGRET in $\gamma$ rays. Even though Geminga has been one of the most intensively studied isolated neutron star during the last 30 years, its interest remains intact especially at $\gamma$-ray energies, where instruments like the Large Area Telescope (LAT) aboard the Fermi mission will provide an unprecedented view of this pulsars. We will report on the preliminary results obtained on the analysis of the first year of observations. We have been able to do precise timing of Geminga  using solely $\gamma$ rays, producing a timing solution and allowing a deep study of the evolution of the light curve with energy. We have also measured and studied the high-energy cutoff  in the phase-averaged spectrum and produced a detailed study of the spectral evolution with phase.
\end{abstract}

\maketitle

\thispagestyle{fancy}


\section{INTRODUCTION}
In the last 30 years Geminga was for a long time the first and only radio-quiet $\gamma$-ray pulsar detected in the sky. After $Fermi$ launch it is no more alone, for other  $\gamma$-ray pulsars have been discovered without a radio or X-ray counterpart so far. Since its discovery Geminga has been widely and deeply studied, but nowdays its importance is increased since it is the harbinger of a population of radio-quiet neutron stars. 
Geminga was discovered in 1975 as a  $\gamma$-ray source by SAS-2~\cite{fichtel75} but only in 1992 it was identified as a pulsar object thanks to the \textit{ROSAT} discovery of a periodic X-rays emission from this source
\cite{halpern92}. This has allowed a successful search for periodicity in
high-energy $\gamma$ rays with EGRET~\cite{bertsch92}. Recent parallax and proper motion
measurements confirm the earlier results, yielding a distance of 250$^{+120}_{-62}$ pc
and a proper motion of 178.2 $\pm$ 0.4 mas\//yr \cite{faherty07}.
Geminga has a period of 237 ms and a period
derivative of 1.1 $\times$ 10$^{-14}$ s s$^{-1}$; these parameters imply that
it is an old $\gamma$-ray with a characteristic age of 3
$\times$ 10$^{5}$ yr and spin-down luminosity $\dot{E}$ = 3.26
$\times$ 10$^{34}$ erg s$^{-1}$.
The light curve detected by EGRET showed two peaks separated by almost half rotation \cite{mayer94,fierro98}. The detected spectrum was compatible with a simple power law with a photon index of $1.42$ even if there was evidence of a cutoff at $\sim$ 2 GeV, but the limited EGRET statistics did not allow a measurement.
Deep X-ray observations allowed \textit{XMM-Newton} and \textit{Chandra} to map
the neutron star surface as it rotates, bringing into view
different regions contributing different spectral components
\cite{caraveo04,deluca05,jackson05} as well as an arcmin-scale bow-shock feature trailing the
pulsar's motion \cite{caraveo03,deluca06}. Such
a non-thermal diffuse X-ray emission surrounding the pulsar is
produced by high-energy ($\sim$ 10$^{14}$ eV) electrons escaping from the pulsar magnetosphere and interacting with
the interstellar magnetic field.\\
\textit{Fermi} is an excellent instrument for high energy $\gamma$-ray 
pulsar studies, thanks to its outstanding timing capability, really good pointing and energy resolution. 
In this proceeding we present the analysis of the Geminga pulsar based on the statistics collected during the first year of operations of the \textit{Fermi} mission. More results on the Geminga observations by $Fermi$ can be found in \cite{abdo09a}.
\section{GAMMA-RAY OBSERVATIONS}\label{aba:sec2}
\textit{Fermi}-LAT is a pair conversion telescope sensitive to $\gamma$ rays from 20MeV to $>$300GeV. It is composed by a silicon tracker, a CsI hodoscopic calorimeter and an Anticoincidence detector to reject the charged particles background~\cite{atwood09}.
The LAT has a large effective area ($\sim$8000 cm$^{2}$ on-axis) and thanks to its large field of view ($\sim$ 2.4 sr), it covers the entire sky every 2 orbits ($\sim$ 3 h). These characteristic lead to an increment of a factor $\sim$ 30 in sensitivity with respect to its predecessor EGRET.
The data used are of particular importance since cover the full first year of the $Fermi$ mission. The collected data range from \emph{Launch and Early Operations} (L$\&$EO) (covering $\sim$ two months after 2008 June 25, when the LAT was operated in pointing and scanning mode for tuning purposes) to the first year of nominal operations (namely 2009 June 15). 
Only for the spectral analysis we selected a subsample of data, covering observations in scanning mode under nominal configuration from 2008 August 4 to 2009 June 15.
We excluded observations when Geminga was viewed at zenith angles $>$ 105$^\circ$~to the instrument axis where Earth's albedo $\gamma$ rays give excessive background contamination. We also have taken care of excluding time intervals when the Region Of Interest (ROI) intersects the Earth's albedo region.
\section{RESULTS}
\subsection{Pulsed profile}\label{aba:sec3}
\begin{figure*}[t]
\centering
\includegraphics[width=135mm]{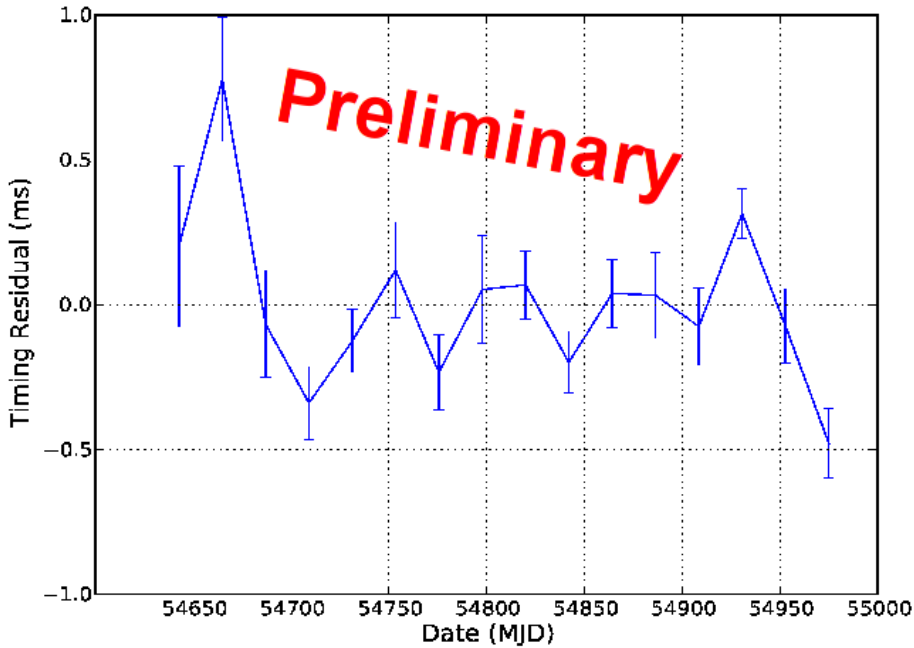}
\caption{Timing solution of Geminga obtained solely from $\gamma$ rays.(The Y-scale is chosen to highlight the timing noise residuals, that corresponds to only ~0.001 in phase)} \label{fig:fig1_gemingarms}
\end{figure*}

The timing solution used to build the light curve has been obtained using only the LAT data, thanks to its excellent timing  accuracy (better than 1$\mu$s). We have used TEMPO2~\cite{hobbs06} to generate a pulse profile and then we have fitted the results with the timing solution model: the residuals have an RMS of 251$\mu$s, as shown in Figure~\ref{fig:fig1_gemingarms}.
Like all the pair conversion telescopes also the \textit{Fermi}-LAT has a Point Spread Function (PSF) that strongly depend with energy. We have chosen to select photons coming from Geminga with an angle $\vartheta < $max[1.6$^\circ$-3$^\circ$ log$_{10}$(E$_{GeV}$), 1.3$^\circ$], that maximize the pulsed signal over the background. 
\begin{figure*}[t]
\centering
\includegraphics[width=135mm]{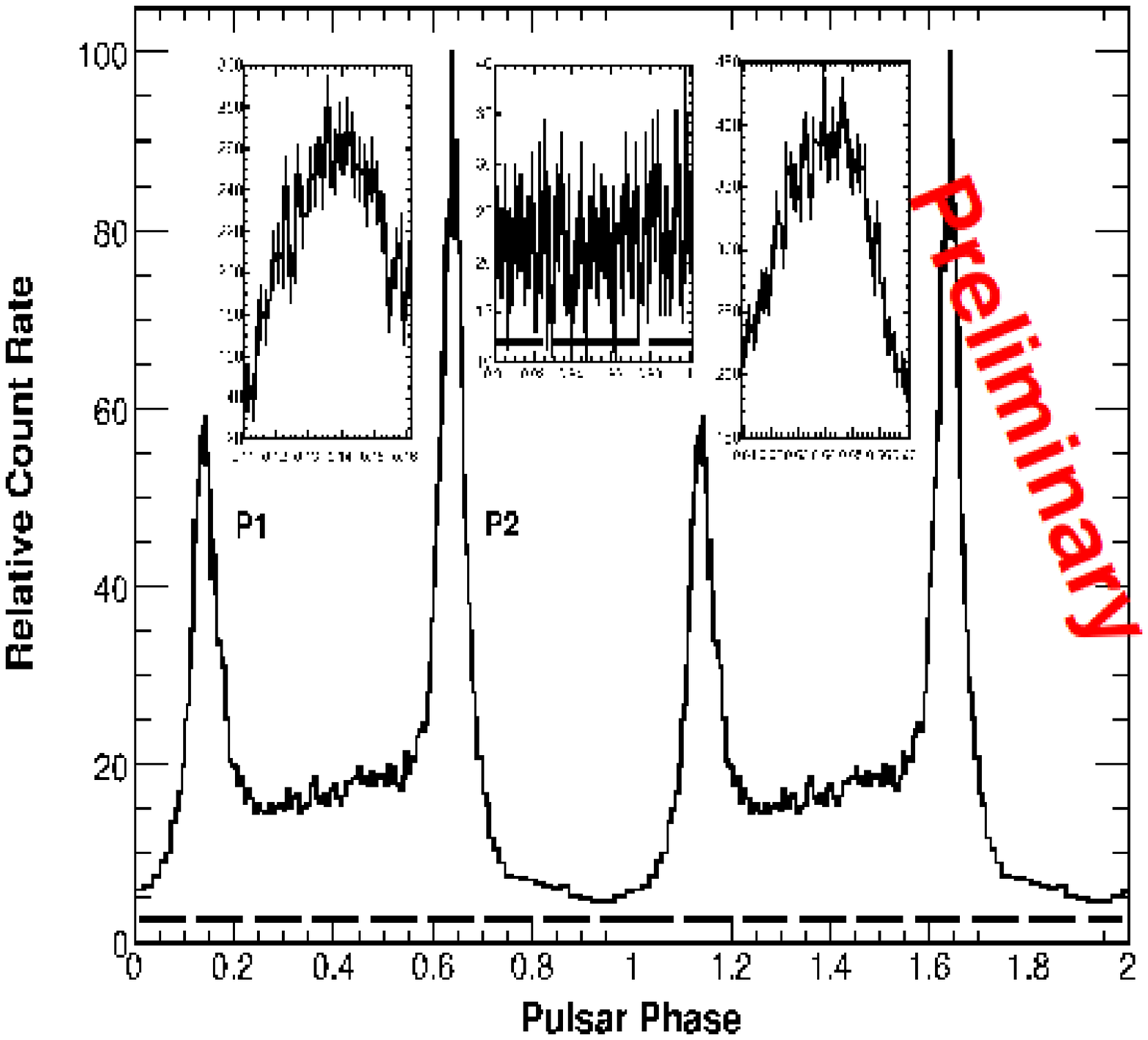}
\caption{Light curve of Geminga above 0.1 GeV. The profile has been plotted in fixed-width phase bins of 400 counts each to highlight fine details. The insets show details of the two peaks and the interpulse} \label{fig:fig2_gemingafulllc}
\end{figure*}
The light curve is shown in Figure \ref{fig:fig2_gemingafulllc}. It has been plotted using a variable-width phase bins, each one containing 400 events, to enhance the fine structure. The photon flux in each phase interval, i.e. the counts divided by phase width, have thus a 1$\sigma$ Poisson statistical error of 5$\%$. The dashed line represents the contribution of the diffuse background, as estimated by selecting photons in the phase interval $\phi$ = 0.9--1.0 from an annular region within 2$^\circ$ and 3$^\circ$~from the source. The total number of pulsed photons is $61219\pm247$ and the backgorund photons are $9821\pm99$. 
Both the peaks have been fitted with an asymmetric Lorentzian function and we obtain the maximum for peak 1 at   
$\phi$ = 0.141 $\pm$ 0.002 and  for peak 2 at $\phi$ = 0.638 $\pm$ 0.003: the separation is  0.497$\pm$0.004. 
\begin{figure*}[t]
\centering
\includegraphics[width=134mm]{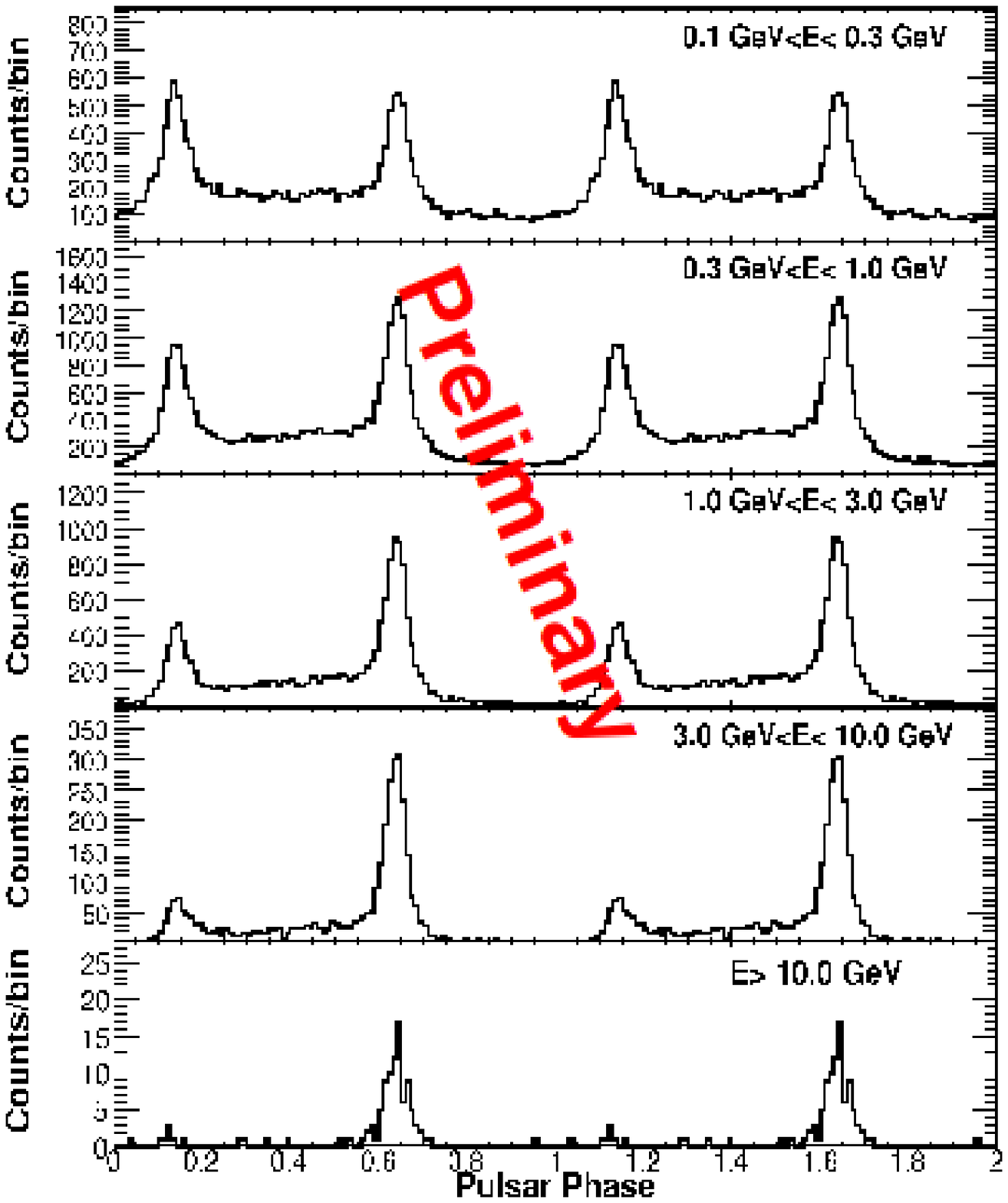}
\caption{Geminga light curves in five energy ranges (0.1--0.3 GeV, 0.3--1 GeV, 1--3 GeV, 3--10 GeV, $>$ 10 GeV). Each light curve is shown over two pulse periods and contains 100 bins/period.} \label{fig:fig3_gemingaedep}
\end{figure*}
We have also built light curves in different energy bands (Figure \ref{fig:fig3_gemingaedep}) and fitted the peaks in each of them. The first trend that can be seen is the first peak is decreasing at higher energies, confirming a decreasing trend in the ratio P1/P2 as observed in Crab, Vela and PSR B1951+32 $\gamma$-ray pulsars by EGRET \cite{thompson04} and now confirmed for many pulsars, including the Vela \cite{abdo09a} and the Crab pulsars \cite{abdo09c} by \emph{Fermi} LAT. We observed pulsed photons up to $\sim$ 18 GeV.\\
Moreover,  both peaks narrow with increasing energy. The decreasing trend in pulse width of P1 and P2 is nearly identical. The emission of the ``first interpeak'' between P1 and P2 is significantly detected up to 10 GeV, emission in the ``second interpeak'' region (between 0.9 and 1.0), not detected before, is clearly present at low energies but vanishes above $\sim$ 2 GeV.\\
\subsection{Energy Spectrum and phase-resolved spectroscopy}\label{aba:sec4}
\begin{figure*}[t]
\centering
\includegraphics[width=134mm]{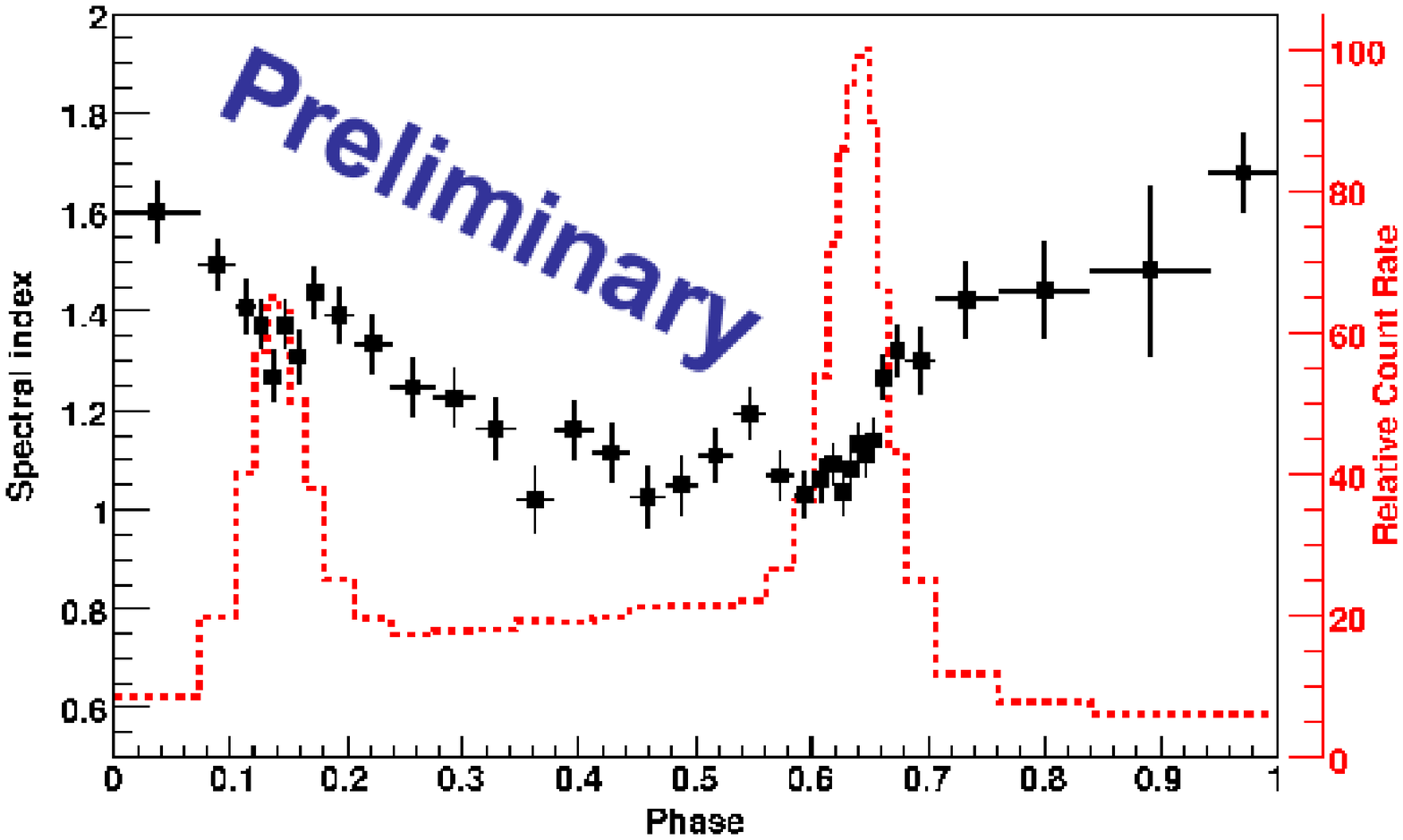}
\includegraphics[width=134mm]{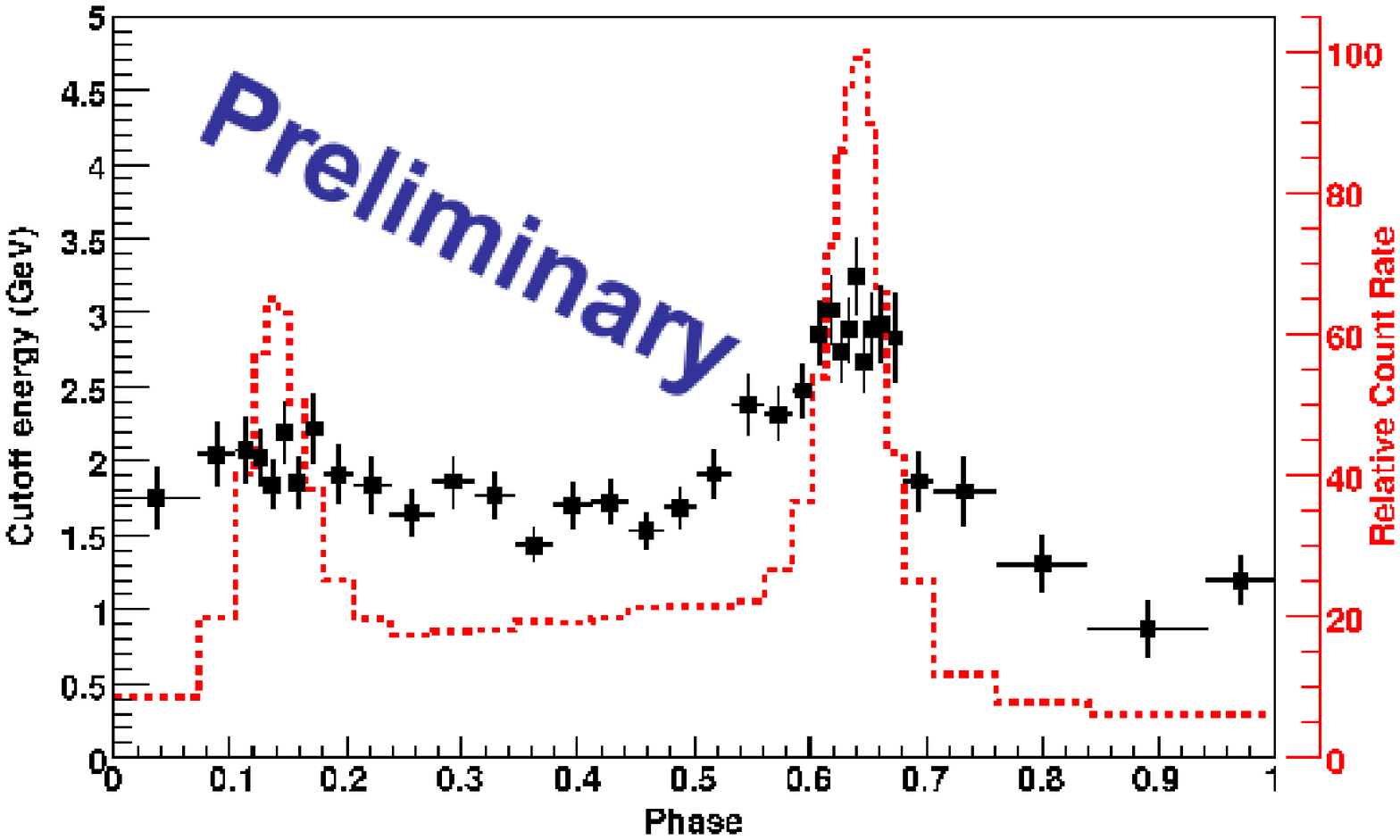}
\caption{Phase evolution of the photon index (\emph{top}) and energy cut-off (\emph{bottom}) above 0.1 GeV as the function of the pulse phase, divided in phase bins each containing 2000 photons. The dashed histogram represents the $Fermi$-LAT light curve above 0.1 GeV in variable-width phase bins of 2000 photons/bin.} \label{fig:fig4_spindex}
\end{figure*}

The spectral analysis is one of the key and most powerful analysis for pulsars. To perform the spectral analysis we have used the maximum-likelihood estimator \emph{gtlike}, that is part of the standard \textit{Fermi} Science Tools provided by the FSSC\footnote{http://fermi.gsfc.nasa.gov/ssc/data/analysis/scitools/overview.html}. The fit was performed using a region of the sky 15$^\circ$~around the pulsar position between 0.1 and 100 GeV.
To properly fit the source we have taken into account all the nearby source in a 20$^\circ$ around Geminga, the galactic diffuse emission based on GALPROP model~\cite{strong04a} and the isotropic residuals emission based on a model derived from a fit of LAT data at high latitude.\\
The function that we have used to fit the spectral emission is a power law with exponential cutoff in the form:
\begin{small}
\begin{equation}
\frac{dN}{dE} = N_{0} E^{-\Gamma} \exp \left(- \frac{E}{E_{0}}\right)  \mbox{cm$^{-2}$s$^{-1}$GeV$^{-1}$}
\label{eq_plec}
\end{equation}
\end{small}
The fit gives the following spectral parameters $N_{0}$ = (1.19 $\pm$ 0.01 $\pm$ 0.08) $\times$ 10$^{-6}$ cm$^{-2}$ s$^{-1}$ GeV$^{-1}$, $\Gamma$ = (1.30 $\pm$ 0.01 $\pm$ 0.05) and $E_{0}$ = (2.47 $\pm$ 0.04 $\pm$ 0.19) GeV. First error is statistical and the second is systematic. The integral flux above 0.1 GeV is (4.15 $\pm$ 0.02 $\pm$ 0.39) $\times$ 10$^{-6}$ 
cm$^{-2}$ s$^{-1}$ and the corresponding energy flux is (2.57 $\pm$ 0.01 $\pm$ 0.32) $\times$ 10$^{-3}$ MeV cm$^{-2}$ s$^{-1}$.
We have also fitted the data with a simple power law and with a power law with a super exponential cutoff with the super exponent fixed at 2, and we have rejected both at more then 5$\sigma$ level.\\
Thanks to the huge number of photons collected we were able also to perform phase-resolved spectroscopy. We divided the period in 36 phase bins, each containing 2000 counts, according to the energy-dependent cut presented in Sec. \ref{aba:sec3}. We have found that the emission is always well described by a power law with exponential cutoff, even if with different spectral parameters. This means that the emission in the two peaks is different, as expected from the study of the light curve with energies, and that the pulsar emission is clearly visible in both the interpulse and in the off-pulse region. The phase evolution of the photon index and the cutoff energy (Figure \ref{fig:fig4_spindex}) show a clear variation across the whole rotation. One of the main features is that the ``second interpeak'' in the range $\phi$ $\sim$ 0.9--1.0 shows an exponential cutoff, indicating that the pulsar emission extends over the whole rotation.
\section{DISCUSSION}
One of the key feature of Geminga is the absence of radio emission, that clearly favor the outer magnetospheric emission models. 
The current evidence against low-altitude emission in $\gamma$-ray pulsars \cite{abdo09a} can also be supplemented by constraints on a separate physical origin. The maximum energy of the pulsed photons observed must lie below the $\gamma$--B pair production mechanism threshold, providing a lower bound to the altitude of the $\gamma$-ray emission $r_{min}$. According to \cite{baring04}, the
highest energy of pulsed photons observed from Geminga (E$\sim$18 GeV), points toward a limit of $r_{min}\ge2.7$ stellar radii, a value clearly precluding emission very near the stellar surface, adding to the advocacy for a slot gap or outer gap acceleration locale for the emission in this pulsar.\\
Following the atlas of $\gamma$-ray light curves compiled by \cite{watters09}, we can use Geminga’s light curve to estimate, for each model, the star’s emission parameters, namely the Earth viewing angle $\zeta_{E}$ with respect to the neutron star spin axis, and the inclination angle $\alpha$ between the star’s magnetic and rotation axes. For Slot Gap/Two Pole Caustic model \cite{muslimov04,dyks03} we obtain $\alpha$ in the range of $30-80,90$ degrees and $\zeta$ in the range $90,55-80$ degrees, while for Outer Gap \cite{romani96} models we obtain $\alpha$ in the range of $10-25$ degrees and $\zeta$ $\sim$ $85$ degrees.\\
The total luminosity radiated by the pulsar is then given by $L_\gamma=4\pi f_\Omega F_{obs}D^2$ where $F_{obs}$ is the observed phase-averaged energy flux over $0.1$ GeV and $D=250_{+120}^{-62}$ pc is the pulsar distance \cite{faherty07}. The beaming factor $f_{\Omega}$ can be derived from \cite{watters09} and points to a value of $0.7-0.9,0.6-0.8$ for Slot Gap and $0.1-0.15$ for Outer Gap. The estimated averaged luminosity is then $L_{\gamma}$=3.1$\times$10$^{34}f_\Omega$ erg s$^{-1}$, yielding a $\gamma$-ray efficiency $\eta_{\gamma}=\frac{L_\gamma}{\dot{E}}$ = 0.15$f_{\Omega}$ ($d$/100pc)$^{2}$. Altough the uncertainty in the distance limit the conclusions, the phase-resolved spectral variation observed can be predicted by caustics models.
\section{SUMMARY}
We presented the analysis of Geminga based on data collected during the first year of operations of $Fermi$. We were able to obtain a timing solution based only on $\gamma$ rays and to produce a high-detailed light curve, that shows clear energy-dependent variation of the profile. Although the phase-averaged spectrum is consistent with a power law with exponential cut-off, the phase-resolved analysis showed a much richer picture of different spectral components intervening at different rotational phases. The phase-resolved analysis has also allowed the detection of the ``second interpeak'' emission indicating a pulsar emission extending over all phases. Our results favor the outer magnetospheric origin for the $\gamma$-ray emission and future improvements in estimating the distance of Geminga will help to better strenghten the conclusions and constraining outer magnetospheric models.

\bigskip 
\begin{acknowledgments}
The \textit{Fermi} LAT Collaboration acknowledges generous ongoing support
from a number of agencies and institutes that have supported both the
development and the operation of the LAT as well as scientific data analysis.
These include the National Aeronautics and Space Administration and the
Department of Energy in the United States, the Commissariat \`a l'Energie Atomique
and the Centre National de la Recherche Scientifique / Institut National de Physique
Nucl\'eaire et de Physique des Particules in France, the Agenzia Spaziale Italiana
and the Istituto Nazionale di Fisica Nucleare in Italy, the Ministry of Education,
Culture, Sports, Science and Technology (MEXT), High Energy Accelerator Research
Organization (KEK) and Japan Aerospace Exploration Agency (JAXA) in Japan, and
the K.~A.~Wallenberg Foundation, the Swedish Research Council and the
Swedish National Space Board in Sweden.
Additional support for science analysis during the operations phase is gratefully
acknowledged from the Istituto Nazionale di Astrofisica in Italy and the Centre National d'\'Etudes Spatiales in France.
\end{acknowledgments}

\bigskip 

\begin{thebibliography}{9}   

\bibitem{abdo09a} Abdo, A.~A., et al.\ 2009a, ApJ, submitted (arXiv:0910.1608)
\bibitem{abdo09b} Abdo, A.~A., et al.\ 2009b, ApJ, submitted (Geminga paper)
\bibitem{abdo09c} Abdo, A.~A., et al.\ 2009c, ApJ, accepted (arXiv:0911.2412)
\bibitem{atwood09} Atwood, W.~B., et al.\ 2009, ApJ, 697, 1071 
\bibitem{baring04} Baring, M.~G.\ 2004, AdSpR, 33, 552
\bibitem{bertsch92} Bertsch, D.~L., et al.\ 1992, Nature, 357, 306
\bibitem{caraveo03} Caraveo, P.~A., Bignami, G.~F., DeLuca, A., Mereghetti, S., Pellizzoni, A., Mignani, R., Tur, A., \& Becker, W.\ 2003, Science, 301, 1345 
\bibitem{caraveo04} Caraveo, P.~A., De Luca, A., Mereghetti, S., Pellizzoni, A., \& Bignami, G.~F.\ 2004, Science, 305, 376  
\bibitem{cheng86} Cheng, K.~S., Ho, C., Ruderman, M.\ 1986, ApJ, 300, 500 
\bibitem{deluca05} De Luca, A., Caraveo, P.~A., Mereghetti, S., Negroni, M., \& Bignami, G.~F.\ 2005, ApJ, 623, 1051 
\bibitem{deluca06} de Luca, A., Caraveo, P.~A., Mattana, F., Pellizzoni, A., \& Bignami, G.~F.\ 2006, A\&A, 445, L9 
\bibitem{dyks03} Dyks, J. and Rudak, B.,\ 2003, ApJ, 598, 1201
\bibitem{faherty07} Faherty, J., Walter, F.~M., \& Anderson, J.\ 2007, Ap\&SS, 308, 225 
\bibitem{fichtel75} Fichtel, C.~E., et al.\ 1975, ApJ, 198, 163 
\bibitem{fierro98} Fierro, J.~M., Michelson, P.~F., Nolan, P.~L., \& Thompson, D.~J.\ 1998, ApJ, 494, 734 
\bibitem{halpern92} Halpern, J.~P., \& Holt, S.~S.\ 1992, Nature, 357, 222 
\bibitem{hobbs06} Hobbs, G.~B., Edwards, R.~T., \& Manchester, R.~N.\ 2006, MNRAS, 369, 655 
\bibitem{jackson05} Jackson, M.~S., \& Halpern, J.~P.\ 2005, ApJ, 633, 1114 
\bibitem{mayer94} Mayer-Hasselwander, H.~A., et al.\ 1994, ApJ,421,276 
\bibitem{muslimov04} Muslimov, A.~G. and Harding, A.~K., ApJ, 588, 430
\bibitem{romani96} Romani, R.~W.,\ 1996, ApJ, 470, 469
\bibitem{strong04a} Strong, A.~W., Moskalenko, I.~V., \& Reimer, O.\ 2004, ApJ, 613, 962 
\bibitem{thompson04} Thompson, D.~J.\ 2004, in Astroph. and Space Sci. 304, Cosmic Gamma-Ray Sources, ed. K. S. Cheng \& G. E. Romero (Kluwer: Dordrecht), 149
\bibitem{watters09} Watters, K.~P. and Romani, R.~W. and Weltevrede, P. and Johnston, S.,\ 2009, ApJ, 695, 1289


\end{thebibliography}

\end{document}